# Observation of Majorana zero modes emerged from topological Dirac semimetal states under uniaxial strain

Quanxin Hu[1, 2, 3], Shengshan Qin[4], Yi Peng[2, 3], Yuke Song[5], Wenyao Liu[6, 2, 3], Yiwei Cheng[3, 7], Renjie Zhang[1, 2, 3], Yudong Hu[1], Chengnuo Meng[7], Yaobo Huang[8], Jin Li[5], Changqing Jin[2, 3], Baiqing Lv[1,9], Jinpeng Xu[10, 11], Hong Ding[1, 11, 12, †]

[1] Tsung-Dao Lee Institute and School of Physics and Astronomy, Shanghai Jiao Tong University, Shanghai 201210, China

[2] Beijing National Laboratory for Condensed Matter Physics and Institute of Physics, Chinese Academy of Sciences, Beijing 100190, China

[3] University of Chinese Academy of Sciences, Beijing 100049, China

[4] School of Physics, Beijing Institute of Technology, Beijing 100081, China

[5] Hunan Key Laboratory for Micro-Nano Energy Materials and Device and School of Physics and Optoelectronics, Xiangtan University, Xiangtan 411105, China

[6] Laboratory for Assembly and Spectroscopy of Emergence, Boston College, Chestnut Hill 02215, USA

[7] Shanghai Institute of Applied Physics, Chinese Academy of Sciences, Shanghai 201800, China

[8] Shanghai Synchrotron Radiation Facility, Shanghai Advanced Research Institute, Chinese Academy of Sciences, Shanghai 201204, China

[9] Zhangjiang Institute for Advanced Study, Shanghai Jiao Tong University, Shanghai 200240, China

[10] School of Physics, Shandong University, Jinan 250100, China

[11] Hefei National Laboratory, Hefei 230088, China

[12] New Cornerstone Science Laboratory, Shanghai 201210, China

† Corresponding authors: dingh@sjtu.edu.cn

**Abstract**

The topological properties observed in iron-based superconductors extend our understanding of vortex Majorana quasiparticle excitations in unexpected ways. Vortex Majorana physics has been extensively studied within the context of the topologically protected surface Dirac state. By employing an *in-situ* strain device, we demonstrate that uniaxial strain can generate Majorana zero modes out of the topological Dirac semimetal bulk state in LiFeAs. Uniaxial strain along [100] direction is found to enhance the band renormalization of LiFeAs, effectively reducing the energy separation between the Fermi level and the topological Dirac semimetal state, and breaking $C_4$ symmetry. Using scanning tunneling microscopy, we observe the evolution of vortex bound states in the topological Dirac semimetal state region, accompanied by the emergence of Majorana zero modes and vortex bound states contributed by the bulk band. Our work provides a controllable method for experimentally engineering Majorana physics in iron-based superconductors, and offers valuable insights into the topological Dirac semimetal state with intrinsic s-wave superconductivity.

The orbital multiplicity of iron-based superconductors (FeSCs), which arises from the $3d^6$ configuration of Fe, can give rise to unusual normal metallic state [1–3], exotic superconducting pairing states [4], unquantized vortices [5–7] and topological band inversion [8,9]. Among these phenomena, $p$–$d$ band inversions along the Γ–Z direction was observed by angle-resolved photoemission spectroscopy (ARPES) in a renormalized electronic band dispersion [8,9]. Compelling evidence of vortex Majorana zero modes (MZMs) have been observed in several FeSCs [10–13], offering a promising new platform for observation and manipulation of MZMs. Up to now, all these observed vortex MZMs were originated from the topological surface state which becomes superconducting below the bulk $T_c$, similar to the Fu-Kane model [14], albeit the superconductivity comes from the material itself. However, this exemplar of topological insulator (TI) with connate superconductivity could be oversimplified when applied to topological FeSCs. As schematically shown in Fig. 1(b), both the bulk topological Dirac semimetal (TDS) states and the helical topological insulating (TI) surface states coexist in many FeSCs [15]. While the vortex topological features in the TI region have been well studied, their counterparts in topological semimetals, e.g., topological Dirac semimetal states have yet to be experimentally explored. Recently, evidence of MZMs in the TDS region have been reported in impurity or natural strain assisted vortices [16,17]. However, these results neither exclude the influence of disorder nor determine the position of the Fermi level.

LiFeAs, a stoichiometric tetragonal iron pnictide exhibiting superconductivity below 18K without intrinsic magnetic or nematic order at low temperatures, has recently garnered significant attention due to its intricate topological band structure, charge-neutral cleavage surface, and small energy separation $\delta_{soc}$ between the TDS and TI states. The energy separation $\delta_{soc}$ can be as small as ~15 meV in LiFeAs. Because of the similar energy scales of the Fermi energy $E_F$, energy separation $\delta_{soc}$ and superconductivity $\Delta$, various nontrivial topological phases and vortex-bound states are predicted based on different positions of the Fermi level [18–20]. Therefore, its vortex topological phase transitions can be manipulated by tuning the chemical potential. Physical pressure, particularly symmetry-specific uniaxial strain, can alter the system's symmetry and significantly change the lattice geometry. This, in turn, affects electron hopping between atoms and modifies the degree of electron correlations, making it especially effective in tuning the properties including chemical potential, superconductivity and topological phase in FeSCs [21–28]. In-plane strain in LiFeAs has been shown to induce phenomena such as wrinkles [29], smectic electronic order [30] and biaxial charge density wave (CDW) [17]. Additionally, scanning tunneling microscopy/ spectroscopy (STM/S) studies have identified signatures of MZMs, which are absent on the naturally clean surface but can emerge in vortices under the influence of the [110]-directional strain which moves the Dirac surface state closer to the Fermi level [31]. However, the vortex topological physics of the bulk TDS state have yet to be experimentally explored. In this work, employing *in-situ* strain ARPES measurements, we observe that uniaxial strain along the [100] direction can serve as a tuning knob to enhance the quasiparticle mass in LiFeAs, effectively reducing the energy separation between the TDS Dirac nodes and the Fermi level. Furthermore, *in-situ* strain STM measurements reveal signatures of vortex topological phase transition, characterized by the presence of MZMs emerged from the bulk TDS states under the breaking of $C_4$ symmetry.

As illustrated in Fig. 1(a), the majority of the exposed cleavage surface of LiFeAs consists of a uniformly ordered Li lattice, interspersed with a sparse distribution of defects (Fig. 1(c)). Uniaxial strain along the Li-Li bond direction, corresponding to the [100] direction, was applied using a custom-built sample holder. This holder is capable of continuously exerting mechanical pressure on the mounted sample, with the strain magnitude controlled by rotating a driving screw [31]. To evaluate the impact of the uniaxial strain, we conducted atomically-resolved topographic image measurements of the pristine sample and the sample under [100]-directional strain at 300 mK with identical scanning speeds to minimize thermal drift and piezoelectric creep. The scanning characteristics of the piezoelectric scanner tube in our STM apparatus were meticulously calibrated. The lattice constant of pristine LiFeAs was measured as a =3.75, b = 3.77 Å, exhibiting four-fold ($C_4$) symmetry (Figs. 1(c) and (e)). Application of strain along the [100] direction alters the lattice constants to a = 3.84 Å and b = 3.68 Å, thus breaking the $C_4$ symmetry (Figs. 1(d) and (f)). The lattice length is compressed along the direction of the applied strain, while is increased along the perpendicular direction, as illustrated in Fig. 1(g). We found that the uniaxial strain along the [100] direction not only alters the lattice constants and symmetry, but also affects the superconducting properties. We measured the tunneling spectra of both pristine sample and the sample under [100]-directional strain. Two superconducting gaps are distinctly observed. Square and triangular markers represent the experimental data, while solid lines correspond to the fitting results using the two-gap Dynes equation [30] (Fig. 1(h)) (see Supplementary materials (SM) for details). It is evident that the uniaxial strain along the [100] direction slightly reduces both the outer (red) and inner (blue) superconducting gaps, consistent with previous local-strain STM/S results [29].

After verifying the functionality of the custom-built sample holder, we performed comprehensive ARPES measurements on strained LiFeAs samples to investigate their Fermi surface (FS) and renormalized electronic band dispersion using the same strained sample holder. The band structure of LiFeAs consists two hole-like FSs at the center of the Brillouin zone (BZ) ($\Gamma$) and two electron-like FSs at the corner of the BZ (M) (Fig. 2(a)). Consistent with atomically resolved topographic imaging, applying strain along the [100] direction deforms the $d_{xy}$ ($\alpha$) Fermi surface pocket from a square shape (Fig. 2(b)) to a rectangular one (Fig. 2(c)). This deformation reduces the system's symmetry from $C_4$ to $C_2$. Besides changing the FS morphology, the strain modifies the electronic band dispersion. The electronic band structure of LiFeAs along the $\overline{\Gamma}$-$\overline{X}$ direction (red dashed line in Fig. 2(a)) is shown in Figs. 2(d) and (e). The left panel shows the band structure of unstrained LiFeAs, where the TDS state is above $E_F$ (purple circles). The middle and right panels display ARPES intensity plots under progressively increasing strain. To quantitatively analyze the electronic band dispersion across different strain conditions, we performed parabolic fitting on the momentum distribution curve (MDC) peaks. The dashed lines represent the fitting results for the three hole-like bands, $\alpha$ ($d_{xy}$), $\beta$ ($d_{yz}$), and $\gamma$ ($d_{xz}$). From this analysis, we extracted the effective mass and band tops of the $\alpha$ and $\beta$ bands as the strain increased (Fig. 2(f)). It is evident that, with increasing strain along the [100] direction, the effective masses of the $\alpha$ and $\beta$ bands progressively increase, while their band tops gradually decrease (Fig. 2(g)). This indicates that strain along the [100] direction enhances electron correlations, thereby increasing the band renormalization. Moreover, when the strain exceeds a certain threshold (right panels in Figs. 2(d) and (e)), the $\beta$ band sinks below the Fermi level, shifting downward by 7 meV relative to the Fermi level. The ARPES spectrum also reveals a second Dirac cone originating from the TDS states (right panels in Figs. 2(d) and (e)).

Motivated by our ARPES results, we conducted STM/S experiments on LiFeAs subjected to uniaxial strain along the [100] direction to investigate vortex-bound states under an external magnetic field. Notably, the same LiFeAs sample was used for all STM measurements. The incremental strain was applied by gradually adjusting the driving screw. Figures 3(a–e) depict zero-bias conductance (ZBC) maps around a vortex in LiFeAs under different strain strengths, with corresponding topographies provided in the insets. To eliminate any unpredictable effects from impurities, we focused on impurity-free vortices distributed on a homogeneous surface. For unstrained LiFeAs (labeled as "$P_0$" in Fig. 3(a)), the ZBC map exhibits a star-like shape with tails extending along the Li-Li directions [32]. As strain along the [100] direction increased progressively ("$P_1$ - $P_4$"), the vortex core gradually lost its initial $C_4$ symmetry. The initially squared flux vortices gradually elongated into rhombic shapes along the diagonal direction, reducing the system's symmetry from $C_4$ to $C_2$. Notably, this deformation of the vortex core shape is consistent with the observed distortion of the atomic lattice and the deformation of the Fermi surface. Moreover, the vortex distortion serves as direct evidence of both the presence and orientation of the applied strain.

Simultaneously, the vortex-bound states evolved with increasing strain. In unstrained LiFeAs (labeled as "$P_0$" in Fig. 3(f)), consistent with previous observations [16], two sets of vortex-bound states at non-zero energies were identified: the set of dispersive states associated with the $\alpha$ band (indicated by the pink dashed lines with arrow) and the set of discrete states associated with the $\beta$ band (marked by the cyan dashed lines with arrow). Upon applying strain along the [100] direction (referred to as "$P_1$"), the dispersive states related to the $\alpha$ band began to discretize (Fig. 3(g)), consistent with the observed reduction in the band top of the $\alpha$ band (Fig. 2(g)). As the strain increased to "$P_2$" (Fig. 3h), the dispersive states associated with the $\alpha$ band gradually transitioned into two sets of discrete levels (indicated by the pink dashed lines with arrow $\alpha_1$ and $\alpha_2$). Further increasing the strain to "$P_3$" led to the full development of these two discrete level sets associated with the $\alpha$ band, accompanied by the emergence of a zero-energy in-gap state (Fig. 3(i)). When the strain reached "$P_4$" (Fig. 3(j)), a pronounced zero-bias conductance peak (ZBCP) appeared within most impurity-free strained vortices, alongside several vortex bound states (VBSs) contributed by the $\alpha$ and $\beta$ bands (see Fig. S3 for more details). The intensity of these ZBCPs gradually diminished to zero without splitting as the distance from the vortex center increased. Moreover, with the gradual increase of strain, the discrete levels associated with the $\beta$ band remained largely unchanged. Figure 4a displays the d$I$/d$V$ spectra measured at the vortex centers from Figs. 3(f-j). The cyan dashed lines indicate the two discrete vortex bound states (VBSs) associated with the $\beta$ band, while the pink dashed lines correspond to those linked to the $\alpha$ band. Based on these results, we can construct a physical picture wherein a vortex phase transition occurs with the gradual increase of strain. This transition is characterized by the emergence of a pronounced ZBCP, accompanied by the

VBSs associated with the bulk band. To ensure the reproducibility of our results, we conducted repeated STM/S measurements on different vortices across multiple samples (Fig. S3). Our observations indicate that ZBCPs consistently emerge from most impurity-free vortices under various magnetic fields, provided that the [100]-direction strain is applied to LiFeAs within an appropriate strength range. Statistically, nearly 100% of the impurity-free vortices measured under strained conditions exhibit distinct ZBCPs and VBSs originating from the bulk band (Fig. S4).

The observed ZBCP is reminiscent of the Majorana zero mode (MZM), which have been observed in other known topological FeSCs [10–13,31,33,34]. The energy positions of the discrete VBSs were extracted by fitting the measured d$I$/d$V$ spectra. In addition to the VBSs originating from the bulk band (Fig. S5), five discrete vortex core states ($L_{-1}$ to $L_3$) were identified (Fig. S6(a)), which closely follow an integer-quantized sequence (Fig. S6(b)). The energy spacing between these VBSs is $\Delta E_L \approx 0.51$ meV. The energy level of the $L_{-1}$ VBS arising from the $\alpha$ band ($\alpha_2$) and the $L_{-1}$ CdGM state originating from the topological surface state nearly coincide. By decomposing the measured d$I$/d$V$ spectra into multiple Gaussian-like peaks, we precisely determined the energy of the $L_0$ state, which was found to be exactly at zero energy. The robust ZBCP without splitting and the integer-quantized Caroli–de Gennes–Matricon (CdGM) bound states strongly support the existence of MZM.

We now turn to the discussion of the origin of MZMs. LiFeAs exhibits a range of topological properties (Fig. 1(b)). ARPES measurements have revealed the presence of both TI and bulk TDS states near the Fermi level [15]. However, no MZMs have been detected in free vortices of LiFeAs. Several explanations for this paradox have been proposed. One hypothesis suggests that the Fermi level of pristine LiFeAs is situated within the band bending region of the TI surface state [16,31]. Consequently, two MZMs can emerge in the vortex core of pristine LiFeAs within impurity-free regions. These pairs of unprotected MZMs would fuse to form fermionic bound states (left panels in Figs. 4(d) and (e)). As a result, the VBSs in pristine LiFeAs do not exhibit MZM, and their behavior is fully explained by the bulk bands (left panels in Fig. 4(f)). An alternative explanation proposes that the energy separation $\delta_{SOC}$ between the TI and TDS state is too small for the assumption of an infinite limit to be valid. Therefore, the interaction between the TI and TDS states leads to the emergence of multiple competing topological phases. When the $C_4$ symmetry is broken by vortex-line tilting or curving, the hybrid vortex becomes topologically trivial and loses its Majorana characteristics [19].

Regardless of the specific interpretation, the position of the Fermi level is crucial for the emergence of MZMs. As illustrated in Fig. 2, the strain in the [100] direction enhances electron correlations. This leads to a progressive increase in the effective mass of the $\alpha$ and $\beta$ bands, accompanied by a gradual decrease of the band tops, which brings the Fermi level closer to the TDS region (middle panels in Figs. 4(d)). At this stage, the superconducting vortex line exhibits one-dimensional gapless helical Majorana modes (middle panels in Figs. 4(e)) [18,20]. Such a nodal vortex line protected by crystalline symmetry would display a constant local density of states (middle panel in Fig. 4(f)). Moreover, as illustrated in Fig. 1 and Fig. 2, uniaxial strain also breaks the system's $C_4$ symmetry. Based on the density functional theory (DFT) calculation (Fig. 4(b)), this perturbation can open a gap at the bulk Dirac points, leading to a transition of the bulk TDS state into a new topological insulator (NTI) state. Consequently, as the Fermi level approaches the NTI state (right panels of Fig. 4(d)), the nodal vortex line transitions into a fully gapped, topologically nontrivial phase, resulting in the emergence of MZMs at the ends of the vortex line (right panels of Fig. 4(e)).

Another critical phenomenon that warrants explanation is the distinct characteristics of MZMs arising from the breaking of $C_4$ symmetry in the TDS region, compared to those observed in the TI region [10–13,31,33,34]. The VBSs in the TI region are characterized by a non-splitting ZBCP and integer-quantized CdGM bound states, primarily attributed to the TI surface state [11,12,31]. In contrast, as shown in Fig. 3 and Fig. 4(a), the VBSs in the TDS region not only include the integer-quantized CdGM bound states associated with the TI surface state but also incorporate VBSs arising from the bulk band. The most plausible explanation for this phenomenon is that the emergence of the NTI state is induced by uniaxial strain, which acts as a minor perturbation that breaks the rotational symmetry, resulting in a very small topological gap. DFT calculations indicate that when the strain reaches approximately ~2%, the corresponding topological gap for the NTI state is only about one-third of that in the original TI state (Fig. S2(b)). Theoretical simulations (Fig. S7) suggest that, under these conditions, the MZMs at the ends of the vortex line are not confined to the surface. Instead, they exhibit non-negligible distribution along the z-axis, extending along the vortex line (right panels of Fig. 4(e)). This distribution allows contributions from the bulk bands to mix with the surface measurements. A model calculation of the VBSs, incorporating both the NTI state and bulk bands ($\alpha$) (Fig. S8), is consistent with the experimental results (Fig. 4(c)). Specifically, the $L_{-1}$(NTI) level VBSs contributed by the NTI state and bulk band ($\alpha_2$) are located at nearly the same energy.

In summary, utilizing our custom-built strain device, we found that uniaxial strain along the [100] direction significantly enhances electron correlations, shifting the Fermi level from the band-bending region into the TDS region. Simultaneously, uniaxial strain breaks the $C_4$ symmetry. We observed the coexistence of integer-quantized CdGM states with MZMs emerging from the bulk TDS state, alongside VBSs originating from the bulk bands. Our experimental results elucidate the evolution of VBSs and the formation process of MZMs. This not only provides compelling evidence that the observed zero modes are indeed MZMs but also offers valuable experimental insights into the TDS state with connate superconductivity. Our approach introduces additional spatial degrees of freedom for detecting emergent vortex Majorana quasiparticle excitations in topological semimetals and for investigating the manipulation of their topological properties through symmetry-specific uniaxial strain.

**Acknowledgements**

We thank Lingyuan Kong, Jiong Mei, Gang Li, Jianxin Zhong, Xin Liu and Fuchun Zhang for helpful discussions. H. D. acknowledges support from the New Cornerstone Science Foundation (No. 23H010801236), Innovation Program for Quantum Science and Technology (No. 2021ZD0302700). Q. H. acknowledges support from China Postdoctoral Science Foundation (No. GZB20230421). This work is partially supported by the Synergic Extreme Condition User Facility, Beijing, China.

**Author contributions**

H. D. conceived the experiment. Q.H. performed STM measurements and data analysis. Q. H. and W. L. performed ARPES measurements. Additional ARPES data were collected by Q.H. with the assistance of Y. C.. Y. P. synthesized LiFeAs single crystals. S. Q. performed theoretical analysis. Y. S. and J. L. performed DFT calculations. Q.H. and H.D. wrote the manuscript with input from all other authors. H. D. supervised this project.

**Competing interests:** Authors declare that they have no competing interests.

**Data and materials availability:** All data are available in the main text or the supplementary materials.

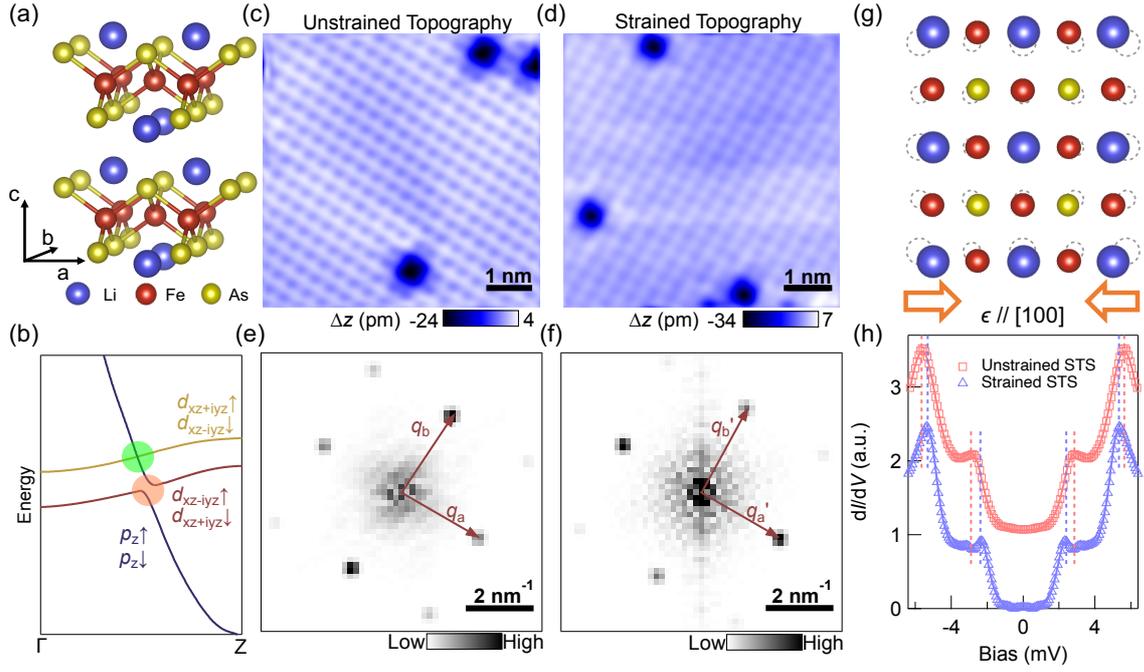

FIG. 1. Crystal structure, topographic image and superconductivity of LiFeAs. (a) Crystal structure of LiFeAs (side view). (b) Band structure of LiFeAs in the normal state. The crossing between ligand $p_z$ orbitals and the iron $d$ bands along the Γ-Z line not only produces topological insulator state (TI) (orange circle), but also bulk topological Dirac state (TDS) (green circle). (c), (d) Atomically-resolved topographic image of pristine LiFeAs (c) ($V$ = -4 mV, $I$ = 260 pA, image size: (6 × 6) nm$^2$) and LiFeAs with compressive strain along [100] direction (d) ($V$ = -4 mV, $I$ = 350 pA, image size: (6 × 6) nm$^2$). (e), (f) The associated Fourier transformations of (c) and (d). (g) Sketch of the crystal deformation under [100]-directional comprehensive strain $\epsilon$. The dashed circles represent the original atomic position. (h) d$I$/d$V$ spectra obtained from unstrained and strained samples. The spectra were measured at 0.3K and have been vertically offset for clarity. Square and triangular markers represent experimental data, while solid lines represent the results fitted using the two-gap Dynes equation.

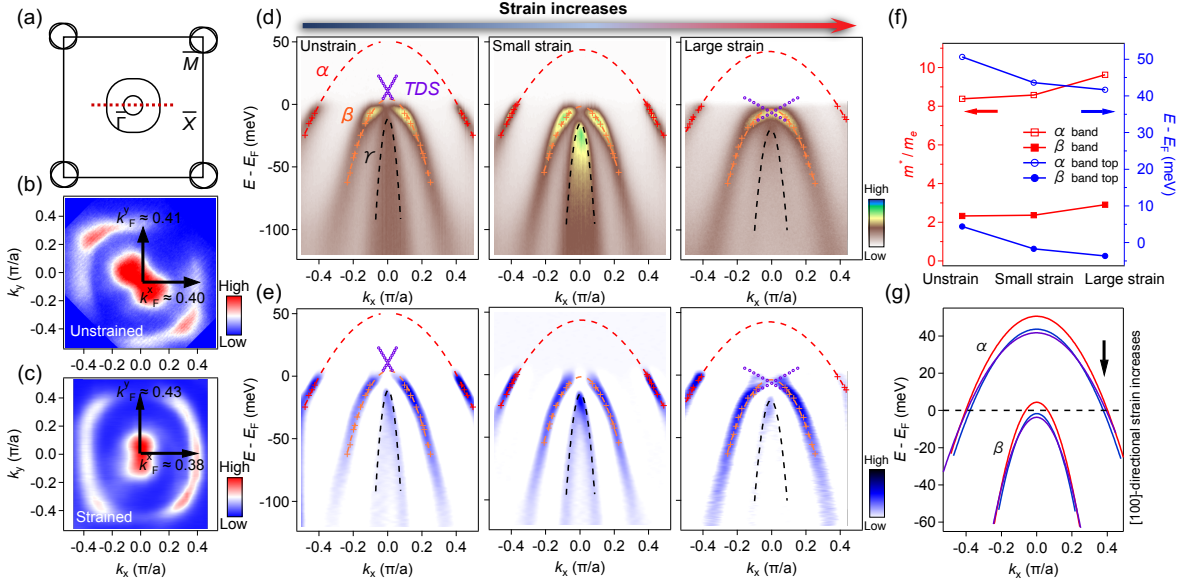

FIG. 2. Orbital selective band renormalization controlled by uniaxial strain. (a) Schematic representation of the surface Brillouin zone (BZ). For pristine LiFeAs, two hole-like Fermi surfaces exist at the $\bar{\Gamma}$ point and two electron-like Fermi surfaces are located at the $\bar{M}$ point. (b), (c) Fermi surface measurements of unstrained and strained LiFeAs were conducted at 20 K. The unstrained LiFeAs was probed with 20 eV incident light in the LH polarization (b), while the strained LiFeAs was measured with a helium lamp ($hv$ = 21.2 eV) (c). The α Fermi surface (FS) pocket undergoes a deformation from square to rectangular under [100]-directional strain. (d) ARPES intensity plot of LiFeAs along $\bar{\Gamma}$ - $\bar{X}$ direction measured at 20 K. The left panel shows the intensity plot measured on unstrained LiFeAs, with 20 eV incident light in the LH polarization. The middle (20 eV incident light in the LH polarization) and right panels (a Helium lamp $hv$ = 21.2 eV) display ARPES intensity plots under progressively increasing strain. The dashed lines are parabolic-type fitting of the momentum distribution curves (MDC) peaks, which extracts the band dispersion. The purple circles represent the TDS band extracted in doped Li(Fe, Co)As [15] (e) Negative second derivative intensity plot of (d). (f) The extracted strain-dependent effective mass and band top of the α and β bands. (g) Fitted band structure of LiFeAs under different strain. As strain increases, the effective mass of the α and β bands increases, while the band top decreases.

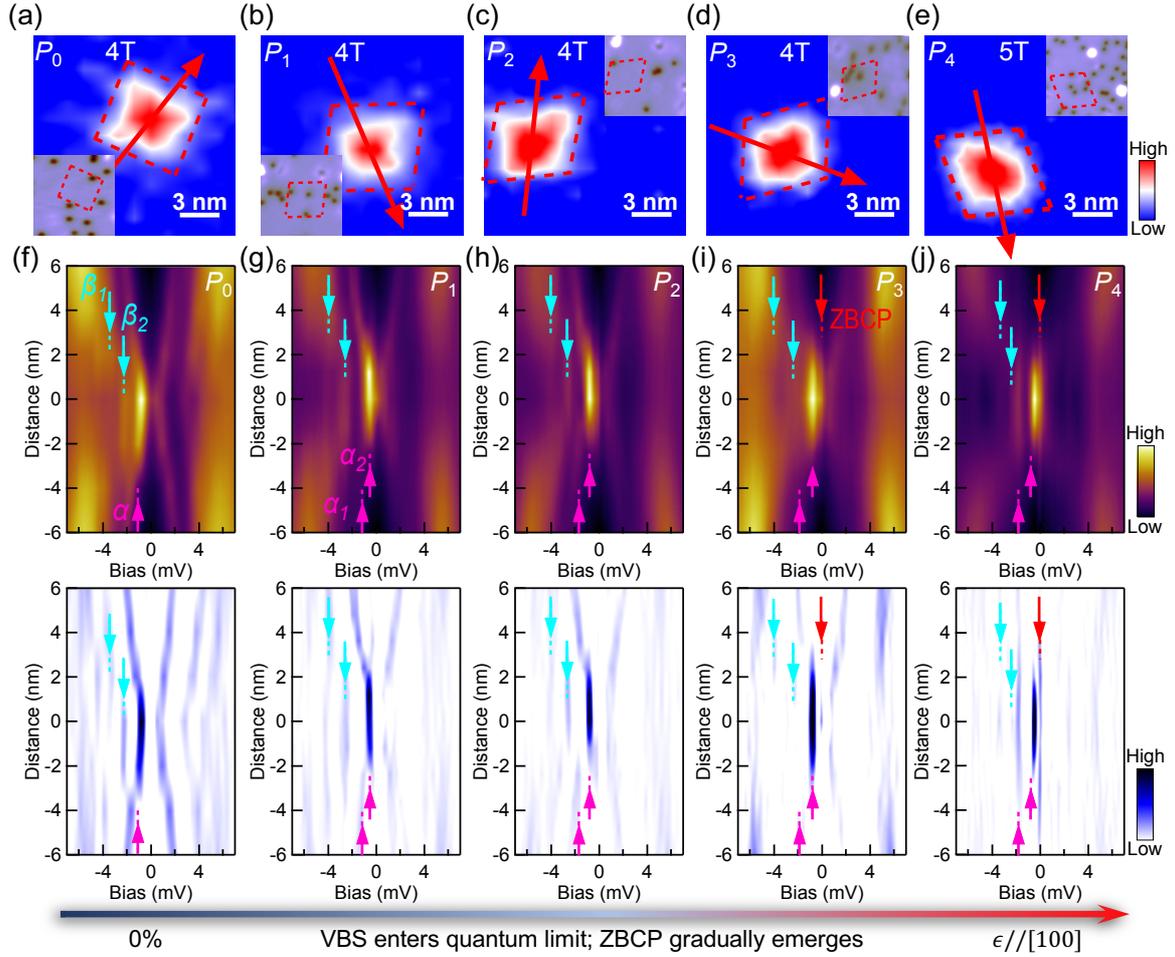

FIG. 3. Evolution of vortex bound states. (a-e) Zero bias conductance (ZBC) maps around vortices under different strain strengths, labeled from "$P_0$" to "$P_4$". The "$P_0$" vortex was measured on unstrained LiFeAs. The uniaxial stress gradually increases from "$P_1$" to "$P_4$". The orange dashed rhombuses indicate the location of the vortices, while the solid arrows represent the direction of the line-cuts. The insets show the topography of the same field of view as the ZBC maps. (f) d$I$/d$V$ spectra measured from unstrained sample ("$P_0$"). The upper panel is the line-cut intensity plot measured along the solid arrow lines shown in (a). The lower panel is the negative second derivative of the upper panel. (g-j), are the same as f, but with the increasing uniaxial strain along [100] direction ("$P_1$ - $P_4$"). The cyan dashed lines with arrows indicate the two discrete vortex bound states at high energy associated with $\beta$ band. The red dashed lines with arrows indicate the vortex bound states associated with $\alpha$ band. The purple dashed line with arrow indicates the ZBCP. With the increasing uniaxial strain along [100], the dispersive vortex bound states ($\alpha$) observed in unstrained LiFeAs ("$P_0$") gradually enter the quantum limit, and an isolated ZBCP appears simultaneously ("$P_2$ - $P_4$").

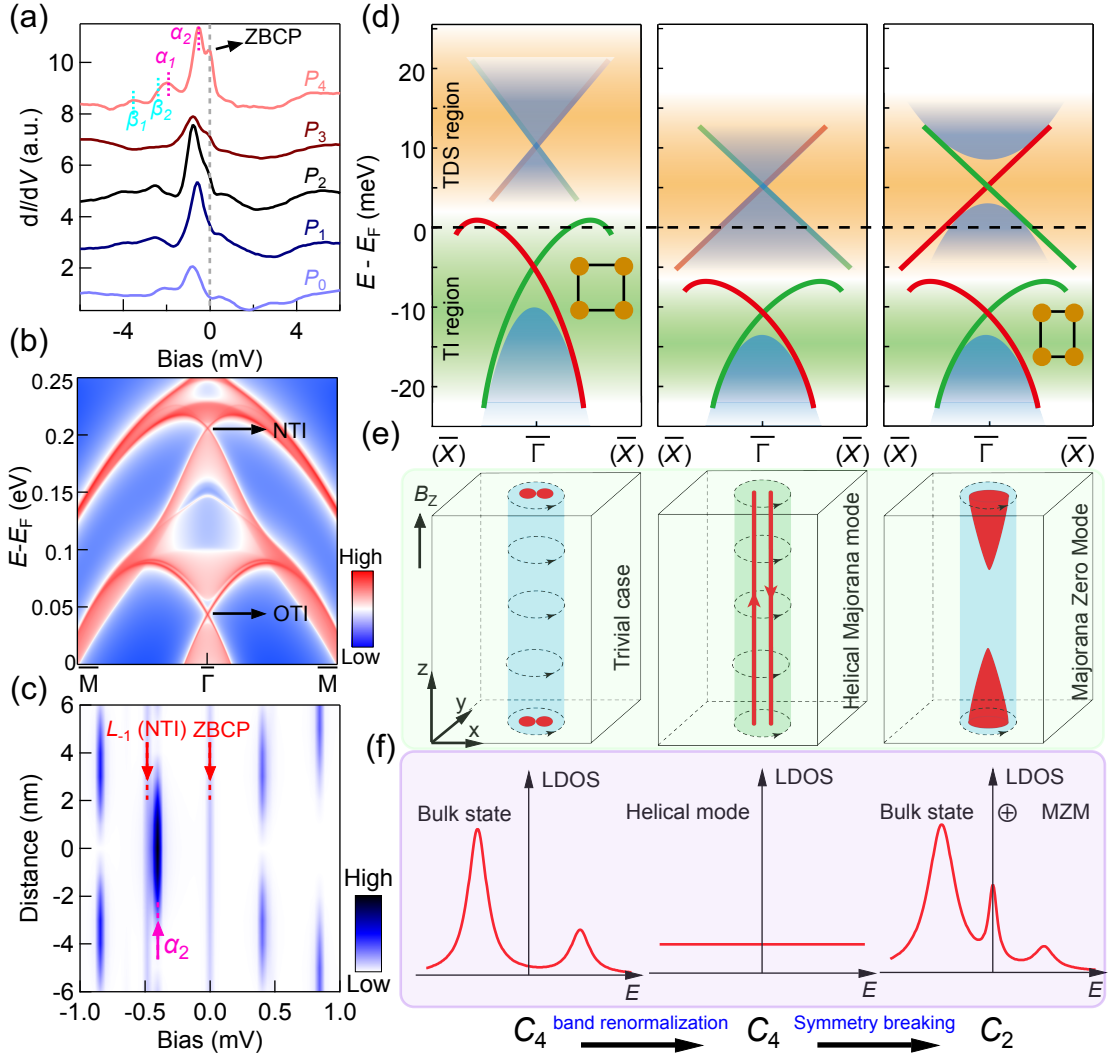

FIG. 4. Coexistence of MZMs and bulk state. (a) d$I$/d$V$ spectra measured at the vortex centers shown in Figs. 3(a-e). The cyan dashed lines indicate the two discrete VBSs corresponding to the $β$ band, while the pink dashed lines represent those associated with the $α$ band. The gray dashed line denotes the zero-bias energy. At the strain level "$P_4$", both the ZBCP and the VBSs associated with the bulk bands emerge simultaneously. (b) DFT calculation results of (001)-surface-projected band structure of LiFeAs under 2% uniaxial strain along [100]-direction. (c) Model calculation of the vortex bound states in LiFeAs, incorporating both the topological surface state and bulk bands. (d) Schematic illustration of the topological band structure of LiFeAs. The left panel shows the band structure of unstrained LiFeAs, where the Fermi level crosses the TI surface state twice. As electron-electron correlations increase, the Fermi level moves closer to the TDS region (middle panel). Uniaxial strain, by breaking $C_4$ symmetry ($C_4 \rightarrow C_2$), disrupts the TDS state and induces the opening of a small topological gap (right panel).

(e) Schematic of corresponding topological phases of the VBSs. VBSs appear as follows: trivial case with a fully gapped state (left panel), helical Majorana mode with a nodal vortex line (middle panel), and Majorana zero mode with non-negligible z-direction distribution (right panel). (f) The LDOS in the case of unstrained LiFeAs (left panel), TDS state with $C_4$ symmetry (middle panel), and $C_4$-symmetry-broken TDS state (right panel).